\documentclass[a4paper]{tCPH2e}

\usepackage{dsfont}
\usepackage{amsmath}
\usepackage{amsfonts}
\usepackage{amssymb}
\usepackage{mathtools}
\usepackage{wrapfig}
\usepackage{graphicx}
\usepackage{chngcntr}
\counterwithout{footnote}{page}

\newcommand{\bra}[1]{\langle #1|}
\newcommand{\ket}[1]{|#1\rangle}
\newcommand{\avg}[1]{\langle #1\rangle}

\def\etal{\mbox{et al.}}

\begin{document}




\title{Gravitational Effects in Quantum Mechanics}

\author{A.~D.~K. Plato$^{\ast}$\thanks{$^\ast$Corresponding author. Email: alexander.plato@imperial.ac.uk
\vspace{6pt}}, C.~N. Hughes and M.~S. Kim\\\vspace{6pt}  {\em{QOLS, Blackett Laboratory, Imperial College London, London SW7 2AZ, UK.}} }

\maketitle
\begin{abstract}
To date, both quantum theory, and Einstein's theory of general relativity have passed every experimental test in their respective regimes. Nevertheless, almost since their inception, there has been debate surrounding whether they should be unified and by now there exists strong theoretical arguments pointing to the necessity of quantising the gravitational field. In recent years, a number of experiments have been proposed which, if successful, should give insight into features at the Planck scale. Here we review some of the motivations, from the perspective of semi-classical arguments, to expect new physical effects at the overlap of quantum theory and general relativity. We conclude with a short introduction to some of the proposals being made to facilitate empirical verification.
\end{abstract}

\section{Introduction}
\label{sec:intro}

Just four years after Heisenberg's and Schr\"odinger's landmark papers, L\'eon Rosenfeld published what would arguably be the first work on quantum gravity \cite{Rosenfeld1930,Rosenfeld1930a}. In the two years prior, Dirac had written down his relativistic wave equation for the electron, and Heisenberg and Pauli had lain the foundations of quantum field theory. With quantisation of the electromagnetic field well under way, it was only natural to include gravity (which at the time, was the only other known field). Yet, despite a few notable efforts, the topic essentially lay dormant until the early 50s -- with the discovery of the neutron, atomic and nuclear physics took centre stage, and the focus was on more easily verifiable theories. In any case, it was already clear that quantising the gravitational field would not be as straightforward as electromagnetism, and given the massive difference in coupling strengths, worries about the instability of atoms due to gravitational radiation could safely be set aside. Indeed, by 1963, Rosenfeld himself questioned whether gravity should even be quantised at all \cite{Rosenfeld1963}.

Nevertheless, in the second half of the century interest in the subject had returned, and by the end of the 1960s both the canonical and covariant lines of research (ultimately leading to loop quantum gravity and string theory) had been defined \cite{Rovelli2000}. Yet despite another 40 years, and a monumental effort, no clear cut theory has yet emerged. This leaves us in the the somewhat uncomfortable position that the two pillars of modern physics still do not seem to work together. By now, numerous books and thousands of papers have been written on the topic, with the main technical and conceptual difficulties well documented (e.g., \cite{Carlip2001}). However, it is clear that one of major impediments is the absence of empirical evidence. In part this is due to a lack of theoretical predictions\footnote{It should be noted of course that both loop quantum gravity and string theory have provided derivations of Hawking radiation, and both predict lengths (or areas) cannot be probed below distances on the order of the Planck length.} but more significant, however, is that most theories predict that the energy scale where quantum gravitational effects become relevant is some 16 orders of magnitude higher than that obtained in the LHC. These energies are so great, that it seems difficult to believe we could hope to attain them in any future particle accelerator.

It was therefore somewhat of a surprise when in 1998, Amelino-Camelia, Ellis, Mavromatos, Nanopoulos, and Sarkar \cite{Amelino-Camelia1998} proposed that observations of photons from gamma ray bursts, originating at cosmological distances, could be used to probe effects at the Planck scale. A common feature appearing in many quantum gravity models is that the velocity at which photons travel through space will depend subtly on their energy. Over cosmological distances, the high energy of the gamma ray bursts should result in a slight, but detectable spread of arrival times. At the same time, Amelino-Camelia also suggested that planned and existing gravitational wave interferometers could be sensitive enough to detect a separate Planck length signature coming from certain models of space-time \cite{Amelino-Camelia1999}. As we shall see, on quite general grounds one can infer that on extremely short scales, space-time may have an inherent ``fuzziness". The idea is that this would then provide an additional source of noise in the interferometer, which for some models was above the level already achieved. These proposals effectively gave rise to a new field -- quantum gravity phenomenology -- and over the last fifteen years the number of papers on the topic has steadily increased. 

More recently, new ideas coupled with advances in the precision measurement of quantum systems have brought about the possibility that Planck scale physics may even be testable in the lab. This prospect is undoubtedly exciting, but also requires some caution - the proposed signatures are not unambiguous, and often lack a concrete underlying model. Nevertheless, this is becoming an increasingly active research area and at the very least, it should encourage some fruitful exchange of ideas. 

The purpose of this review is to give a brief introduction to some of the common features expected to emerge at the overlap of quantum theory and general relativity. Where possible we will try to motivate the ideas on quite general grounds, rather than appealing to a particular theory. We make no claim that this will be in any way comprehensive. A more detailed overview of quantum gravity phenomenology can be found in \cite{Amelino-Camelia2013a}, while for an introduction to the Planck length as a minimum scale see \cite{Garay1995,Hossenfelder2013}. Instead, our aim will be to cover those features most relevant to current experimental proposals, while at the same time giving a flavour of what is being done in that direction.

As the key difficultly is the inaccessibility of the Planck scale, it is worth taking a moment to first expand upon its significance. 

\section{Why the Planck scale?}
In the final section of Max Planck's 1899 paper he noted that taken together, his new constant $h$\footnote{Originally denoted $b$. In the remainder of this review, we shall use the reduced Planck's constant $\hbar=h/2\pi$.} along with the constants $c$ and $G$ could uniquely determine (up to some numerical factor) units of length, mass and time,
\begin{align}
l_p &= \sqrt{\frac{\hbar G}{c^3}} \approx 1.6\times 10^{-35}m \nonumber \\
m_p &=\sqrt{\frac{\hbar c}{G}}\approx 2.2 \times 10^{-8}kg \approx 1.2\times 10^{19}GeV/c^2 \nonumber\\
t_p &= \frac{l_p}{c}  \approx 10^{-43}s. \label{Planckunits}
\end{align}
At that time there was no suggestion that these units had any physical significance, rather Planck was more interested in their universal nature as measures. Indeed, the idea of a fundamental distance would not emerge until much later, when Heisenberg and others were trying to deal with divergences appearing in quantum field theory. In any case, they held the view that this should be much larger, on the order of a femtometer \cite{Heisenberg1938b}. Curiously, it seems the first explicit reference to ``Planck's units" as fundamental quantities wasn't until Wheeler \cite{Misner1957}, though as an unnamed combination, factors of $c,\hbar$ and $G$ as above had appeared as far back as 1936 \cite{Bronstein1936}.

\begin{wrapfigure}{r}{0.5\textwidth}
  \begin{center}
\includegraphics[width=8cm]{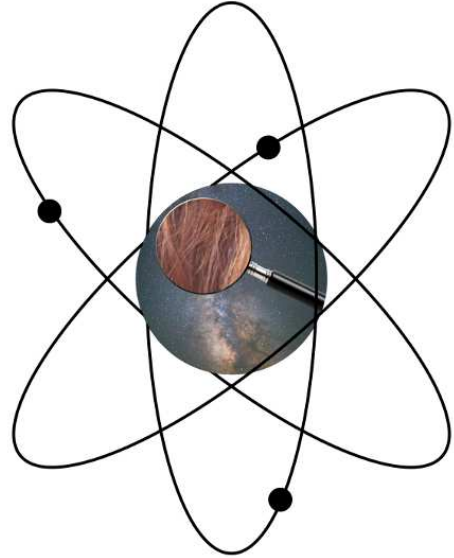}
  \end{center}
  \caption{Visualising the Planck Length: We would need to resolve a single human hair, from within the Milky Way when scaled down to the size of a Hydrogen atom.}
  \label{fig:PlanckLength}
\end{wrapfigure}

To highlight the enormity of the task facing any experiment on the Planck scale, it is worth appreciating just how small the length in (\ref{Planckunits}) is. This is of course, not straight forward. The best electron microscopes can image down to the level of individual atoms, at around $10^{-10}$m. While remarkable, this is still some 25 orders of magnitude larger than $l_p$. To get an idea of how much we need to improve our resolution, consider for a moment a view of the night sky. On a clear evening, far from any stray lights, one should have little problem in discerning the arc of the Milky Way. The amount visible is somewhat restricted (partly down to the angle of view, and partly because interstellar dust obstructs how far out we can actually see), but it is estimated that the diameter is on the order of $100,000$ lightyears $\approx 10^{21}$m. Thus, to see down to the Planck scale, we would need to resolve details on an atom equivalent to $100\mu$m in the size of our own galaxy. For comparison, this is roughly the thickness of a single human hair, Fig. (\ref{fig:PlanckLength}).
\subsection{Minimum Length}
\begin{wrapfigure}{r}{0.5\textwidth}
  \begin{center}
\includegraphics[width=8cm]{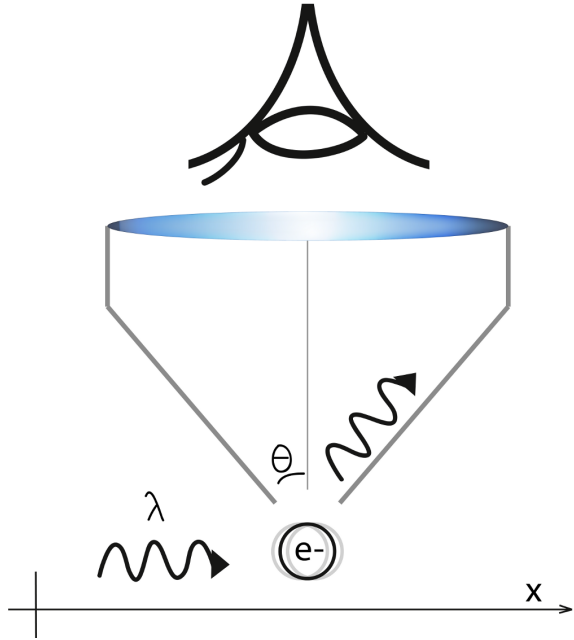}
  \end{center}
    \caption{The Heisenberg Micrscope: A photon with wavelength $\lambda$ scatters off of an electron, before being focused by a lens. The resolution is limited by the diffraction limit, while at the same time the electron is displaced slightly by the scattering. Combining these uncertainties leads to the familiar Heisenberg uncertainty principle.}
  \label{fig:HeisenbergMicroscope}
\end{wrapfigure}

Of course, one can ask what would happen if we actually tried to build a microscope with this kind resolution. A natural starting point is to consider the well known thought experiment used by Heisenberg in his famous uncertainty principle (Fig. \ref{fig:HeisenbergMicroscope}). One imagines that we wish to resolve the position of a charged particle (such as an electron) by scattering a photon into lens. As long as the electron lies somewhere in the focal plane, then we can measure its location (say along the $x$-axis) by looking at the image formed. The accuracy, however, will depend on the the resolving power of the microscope. That is, how big the angular distance between two points must be before they can be distinguished. Even with a perfect lens, the resolution will still be limited by diffraction,
\begin{equation}\label{eq:resolutionlimit}
\Delta x \sim \frac{\lambda}{2\sin{\theta}},
\end{equation}
We can improve our accuracy by probing the particle with shorter wavelengths (and to a lesser extent by increasing the size of the lens). However, Heisenberg pointed out that during the scattering process, the photon transfers some of its momentum to the particle. As we don't know its final direction to better than an angle $\theta$, the uncertainty in the $x$-momentum of the particle must be of order, 
\begin{equation}\label{eq:deltap}
\Delta p_x\sim\frac{h}{\lambda}\sin \theta,
\end{equation}
Multiplying these two uncertainties together, we end up with Heisenberg's original uncertainty principle (HUP),
\begin{equation}\label{microscopeHUP}
\Delta x\Delta p_x \sim \frac{h}{2}.
\end{equation}
The familiar adoption of the standard deviation for the uncertainty was later formalised by Kennard \cite{Kennard1927} which fixed the coefficient as $\hbar/2$ and provided a strict lower bound. This was then generalised to arbitrary symmetric operators, $A$ and $B$, by Robertson \cite{Robertson1929} and Sch\"odinger \cite{Schroedinger1930}\footnote{It should noted that in Heisenberg's original (semi-classical) discussion, the uncertainty arises as a disturbance from the measurement process (referred to as a measurement-disturbance relation), whereas the Kennard-Robertson-Sch\"odinger relation is an intrinsic property of the ensemble of quantum states used to make repeated measurements (preparation uncertainty). The distinction can often lead to controversy, however with a suitable definition of the $\Delta$'s appearing in eq. (\ref{microscopeHUP}), a general quantum version of the measurement-disturbance relation (including the usual coefficients) can be found \cite{Busch2007a,Busch2013}. A discussion of this distinction in the context of quantum gravity is beyond the scope of this review, but some comments can be found in, e.g., \cite{Garay1995} sec. IV.}
\begin{equation}\label{KRSrelation}
(\Delta A)^2(\Delta B)^2 \geq \left|\frac{1}{2}\avg{\{A,B\}} - \avg{A}\avg{B}\right|^2 + \left|\frac{1}{2i}\avg{[A,B]}\right|^2.
\end{equation}

Nevertheless, this relation does not forbid us from minimising the position uncertainty to as small as we like, provided we don't care about the momentum. Indeed, if it did we would no longer be able to define position eigenstates and the very notion of measurement would be ambiguous. Thus, at first glance it seems there should be no problem, in principle, with building a microscope powerful enough to resolve the position of a particle to within the Planck length. 

However, in 1964 Mead argued that this model is incomplete \cite{Mead1964}. In addition to the momentum exchange, the photon also exerts a gravitational attraction on the electron and so its position moves before the measurement can take place. This motion cannot be corrected as we do not know the precise path of the photon. In the simplest treatment, he considered the interaction as Newtonian and noted that in reality the photon doesn't scatter at a point, but rather interacts over some characteristic distance\footnote{For argument's sake, we can imagine that this is on the order of the photon's wavelength.}, $r$. Therefore, over the interaction time, $t=r/c$, the particle feels an acceleration proportional to $E/c^2=h/c\lambda$, 
\[
a\sim \frac{G h}{c \lambda r^2},
\]
picking up a velocity $v=G h /c^2 \lambda r$, and traveling a distance $l\sim Gh/\lambda c^3$. Projecting along the x-axis, the uncertainty due to the gravitational motion is $\Delta x_{grav} \sim Gh \sin \theta/\lambda c^3 \sim l_p^2 \Delta p_x$ (using (\ref{eq:deltap})). Together with the usual Heisenberg uncertainty,  we have (up to some factor of order $1$),
\begin{equation}\label{minL}
\Delta x \gtrsim \max\left(\frac{\hbar}{2\Delta p }, l_p^2 \Delta p \right) \gtrsim l_p.
\end{equation}
Thus we find that now as we try to look closer and closer, gravitational effects begin to become important and eventually we hit a fundamental limit at the Planck scale. This argument is of course quite relaxed and we could question whether this is simply an artifact of using the Newtonian approximation. Mead considered this, and after a careful analysis using general relativity he recovered the same result. However, the notion of a minimum length was not taken seriously, and Mead's paper was largely overlooked until the late 1980s, when results from string scattering led to a similar lower bound \cite{Amati1989}. Since then, a number of new (and almost identical) arguments \cite{Maggiore1993,Scardigli1999,Adler1999} were put forward and it became common to rephrase eq. (\ref{minL}) in the form of a generalised Heisenberg principle (GUP). To do this, we simply assume the two uncertainties add linearly (with the weighting described by some factor, $\alpha$, of the order 1) and then multiply through by $\Delta p$,
\begin{equation}\label{GUP}
\Delta x \Delta p= \frac{\hbar}{2} +  \alpha l_p^2 (\Delta p)^2.
\end{equation}
This relation is often used as the starting point for a variety of minimum length studies, and we will discuss some of its immediate consequences in section \ref{sec:Implications}.

\subsection{Fluctuations of Space-Time}
One can also argue the significance of the Planck scale in another way. It was well known that when the electromagnetic field is quantised, it imposes uncertainties on the measurement of individual field components. By applying the HUP to the position and momentum of a measuring apparatus, Bohr and Rosenfeld found a set of relations which depend on the volume of the device, $V$ and the time over which the field is measured, $T$. For the $E_x$ component this takes the form, 
\[
\Delta E_x \geq \sqrt{\frac{\hbar}{V T}}.
\]
Where the field components should be averaged over $V$ and $T$ to have a well defined meaning. The measuring apparatus should, for all intents and purposes, be a classical device. Assuming the gravitational field should also be quantised, Bronstein \cite{Bronstein1936} quickly applied their reasoning to the weak field limit. However, he noted that unlike electromagnetism, gravity does not allow arbitrarily high ``charge" (i.e., mass) densities to be confined in a small region of space-time. He reasoned the uncertainty of the Christoffel symbol, $\Gamma^{00}_1$ should take on the form,
\[
\Delta \Gamma^{00}_1 \geq \frac{h^{2/3}G^{1/3}}{c^{1/3}\rho^{1/3}V^{2/3}T}.
\]
However, the Schwarzschild radius $\sim G\rho V/c^2$ of the test body must be smaller than the its linear extent $l\sim V^{1/3}$. This means the density must be less than $\rho \sim c^2/GV^{2/3}$ and so, 
\[
\Delta \Gamma^{00}_1 \geq \frac{c}{T} \left(\frac{l_p}{l} \right)^{4/3}.
\]   
Later, Wheeler noted that in analogy with Bohr and Rosenfeld's uncertainty the quantum fluctuations in a typical gravitational potential would be \cite{Wheeler1955} (see also \cite{Peres1960}),
\begin{equation}\label{Wheeleruncertainty}
\Delta g \approx l_p l^{-1},
\end{equation}
These would be negligible so long as the extension in space-time, $l$, was much larger than the Planck length. However, once that scale was reached the metric could vary wildly, and one would need to go beyond the weak limit.

Along similar lines, Salecker and Wigner \cite{Salecker1958} explored the fundamental limits to space-time measurements. They proposed that such distances should be measured by only using clocks, rather than fixed rods. To achieve a given accuracy, one had to then impose certain physical requirements on the time measuring device (see also \cite{Peres1980}), such that the accuracy, $\Delta T$, for measuring a time interval $T=t_2-t_1$, of a quantum clock was proportional to the square root of the length of observation divided by its mass, $m$,
\begin{equation}\label{Salecker-Wigner}
(\Delta T)^2 \gtrsim \frac{\hbar}{mc^2}T,
\end{equation}
or equivalently, for distances $l=cT$, $(\Delta l)^2 \gtrsim (\hbar/mc)l$. This relation has generated a vast literature, and various examples for the construction of such clocks have been put forward. A somewhat heuristic argument for its derivation can be made (\cite{Salecker1958,Karolyhazy1966}) by considering that any microscopic clock must be able to transmit its timings to an external, macroscopic observer (otherwise the clock itself could not be microscopic). The overall uncertainty in $T$ should then be at least as large as the uncertainties $\Delta x(t_1)/c$ and $\Delta x(t_2)/c$ at the two ends of the interval. However, Heisenberg's uncertainty relation implies that at $t_1$ the clock will have an uncertain velocity $\Delta v \geq \hbar/2m \Delta x(t_1)$, so that by end of the interval,
\begin{equation}\label{Kp1}
\Delta x(t_2) \geq \frac{\hbar}{2m \Delta x(t_1)}T.
\end{equation}
The uncertainty in $T$ is then minimised when $\Delta x(t_1)=\Delta x(t_2)$, which when assumed to be approximately equal to $\Delta l =c \Delta T$ recovers (up to a factor of $2$) the bound in eq. (\ref{Salecker-Wigner}). In analogy with the Bohr and Rosenfeld result, the Salecker-Wigner relation introduces a dependence on the parameter valued time of observation. On the other hand, just as in the HUP, $l$ (or $T$) can still be measured to arbitrary accuracy by taking the limit $m\rightarrow \infty$. This parallels the comments made earlier.

K{\'a}rolyh{\'a}zy however, pointed out that at some point an increasingly massive quantum clock will eventually be subject to gravitational effects (\cite{Karolyhazy1966}). In order to continue signaling, the clock must remain larger than its Schwarzschild radius, $R_s=2Gm/c^2$. Thus to minimise the uncertainty, $\Delta l=\Delta x \approx R_s$. Rearranging, K{\'a}rolyh{\'a}zy proposed\footnote{It should be noted that K{\'a}rolyh{\'a}zy originally kept the discussion quite general, and so did not restrict his relation to clocks.},
\begin{equation}\label{Karolyhazyuncertainty}
(\Delta l)^3 \gtrsim \left( \frac{G \hbar}{c^3}\right) l = l l_p^2.
\end{equation}

The relation above is also sometimes referred to as a holographic scaling \cite{Ng2000}. This is because the number of degrees of freedom, thought of as minimal uncertainty cubes (of side $\Delta l$), that one can fit into a cube of volume $l^3$ goes as $(l/\Delta l)^3$. However the holographic principle \cite{Hooft1993} suggests this is bounded by the surface area of the region

\section{Implications of the Planck Scale}\label{sec:Implications}

\subsection{Modified Dispersion Relations}

While a minimum length (at the Planck scale or otherwise) would have its benefits -- for example, as a UV cutoff in quantum field theory -- it was recognised very early on that it should also require a reassessment of Lorentz symmetry. For example, a distance measured in the frame of one observer to be $l_p$ should be contracted by an amount $\sqrt{1-v^2/c^2}$ for another traveling at relative velocity $v$. This immediately seems to suggest that Lorentz invariance must be broken at very small distances. The implications of this type of symmetry breaking were studied in various contexts (and initially with no reference to the gravitational field). For example, Pavlopoulos \cite{Pavlopoulos1967} pointed out that Poincar\'e's derivation of the Lorentz transformations postulated the invariance of the wave equation,
\[
\nabla^2 \Psi - \frac{1}{c^2}\frac{\partial^2 \Psi}{\partial t^2} = 0,
\]
under coordinate transformations. If a minimum length, $l_0$, is also to be introduced as a fundamental invariant (as with $c$), then on dimensional grounds the wave equation can only be modified by adding higher order derivatives. He assumed that such an extension should continue to admit plane wave solutions, $\Psi=\Psi_0e^{i(\omega t -kx)}$, and among other things, reduce to the standard form in the limit $l_0\rightarrow 0$ \cite{Pavlopoulos1967}. A simple example fulfilling these conditions was, 
\[
-l_0^2 \nabla^4 \Psi + \nabla^2 \Psi - \frac{1}{c^2}\frac{\partial^2 \Psi}{\partial t^2} = 0.
\]
The requirement of a plane wave solution then implies a modification of the dispersion relation, 
\begin{equation}
\omega^2 = c^2 k^2(1 + l_0^2 k^2).
\end{equation}
An immediate consequence is that now the group velocity is no longer equal to $c$. For the example above, in the limit that $l_0^2k^2 \gg 1$ then $d\omega/dk \approx c(1+(3/2)k^2l_0^2)$.

Following the results from string theory \cite{Amati1989}, a number of other candidate theories began recovering similar minimal length scales, and by the late 1990s a general picture of a modified energy dispersion relation had emerged. These were typically characterised by a phenomenological modification of the form,
\begin{equation}
E^2 = m^2 c^4 + p^2 c^2 - f(m,p; m_p),
\end{equation}
where $f(E,m,p; l_p)$ is expanded as a power series in $1/m_p$ (with factors of $m$, $p$ and $c$ to preserve dimensions). The most studied case is the linear correction, which has been particularly relevant in observations of high energy gamma ray bursts. Here the existence of the function $f$ means that the photon ($m=0$) group velocity now has an energy dependence $dE/dp \approx c(1-\xi E/E_p)$, where the coefficient $\xi$ depends on the dynamical framework. This results in a difference of arrival times between two photons, with energy difference $\delta E$, emitted simultaneously from a (nearly) non red shifted source at distance $L$ \cite{Amelino-Camelia2009b},
\[
\delta t \approx \xi\frac{\delta E}{m_p}L.
\]

By viewing $l_p$ as a second observer-independent scale, one can also be led to the idea that rather than breaking Lorentz invariance outright, what we need instead is a modification of the Lorentz transforms. This has given rise to so called models of Deformed (or Doubly) special relativity (DSR) \cite{Amelino-Camelia2001a,Kowalski-Glikman2001,Amelino-Camelia2002a,Magueijo2002}. In the DSR framework, one can still find a modified dispersion relation, however to preserve the new two scale notion of relativistic invariance one must modify energy-momentum conservation \cite{Amelino-Camelia2010}. This in turn leads to difficulties in dealing with multi-particle states, commonly referred to as the ``soccer ball problem" \cite{Hossenfelder2014a}.

On the other hand, there exists the possibility that Lorentz symmetry is neither broken or deformed. For example, in loop quantum gravity the minimum length (expressed through an area) appears as the minimal eigenvalue of a quantum observable \cite{Rovelli2003}. A boosted observer instead sees a different probability distribution for measuring one of many allowable areas. Other examples include analogies with condensed matter physics, where space-time is viewed as an emergent phenomena, \cite{Laughlin2003}, or from discrete models, such as causal sets \cite{Dowker2006}. 

\subsection{Modified dynamics and deformed commutators}\label{deformedcommutators}

The modification to the Heisenberg uncertainty relation proposed in eq. (\ref{GUP}) can also be expected to lead to subtle corrections in familiar quantum systems. This can be seen quite easily by repeating a classic estimation for the ground state energy of the Hydrogen atom. Classically, we can write the energy of the bound electron as, 
\[
E\sim \frac{p^2}{2m_e} - \frac{\tilde{e}^2 }{r},
\]
where $\tilde{e^2} \equiv q_e^2/4\pi\epsilon _0$. The usual argument is to note (as we implicitly have above) that any measurement must be at least as large as its associated uncertainty. Therefore, $p \gtrsim \Delta p$ and $1/\Delta r \gtrsim 1/r$. Applying the usual HUP lets us rewrite the above equation in terms of only the $r$ variable, for which a solving $dE/dr=0$ returns a reasonable estimate of the ground state energy. In the case of the GUP, the minimum uncertainty $\Delta p$ is no longer simply inversely proportional to $\Delta x$, but takes the form, 
\begin{equation}\label{deltapGUP}
\Delta p = \frac{\Delta x}{\hbar \beta} \pm \sqrt{\left(\frac{\Delta x}{\hbar \beta}\right)^2+\frac{1}{\beta}}
\end{equation} 
where we have grouped the terms $2\alpha l_p^2/\hbar \equiv \beta$ for convenience. Taking a first order expansion in $\beta$ and repeating the same procedure leads to an additional correction of the order \cite{Bambi2008},
\[
\Delta E_0 \sim \beta m_e^3 e^8 \sim 10^{-48}\text{eV}.
\]
Similar corrections appear for the harmonic oscillator, and indeed for quantum mechanical systems in general. However, without providing the underlying algebraic structure and state space it is difficult to make any concrete statements. To this end, we can recall from the Robertson form of the eq. (\ref{KRSrelation}) that the HUP depends on the commutator $[x,p]$. Therefore, we expect that a new commutation relation must underly the GUP eq. (\ref{GUP}). It is easily checked that the following gives the desired result \cite{Maggiore1993a,Kempf1994,Kempf1995},
\begin{equation}\label{deformedcommutator}
[x,p]=i\hbar(1+\beta p^2).
\end{equation}
Generalisation to 3-dimensions using the Jacobi identity then leads naturally to a non-commutative space-time structure.

Of course, a minimum length implies our state space can no longer contain position eigenstates. This is problematic for defining a Hilbert space representation on position wavefunctions (\cite{Kempf1995}). On the other hand, the momentum uncertainty is not similarly restricted, and so we can instead represent the Heisenberg algebra on momentum space wavefunctions, $\psi(p) \coloneqq \langle p | \psi \rangle$. A representation which achieves this is,
\begin{align}
{p}. \psi(p) &= p \psi (p), \\
{x}. \psi(p) &= i\hbar(1+\beta p^2) \partial_p \psi(p),
\end{align}
but now we must define a new scalar product,
\begin{equation}\label{scalarproduct}
\langle\psi|\phi\rangle=\int^{+\infty}_{-\infty} \frac{dp}{1+\beta p^2} \psi^*(p)\phi(p).
\end{equation}
While position eigenstates are not physical, it turns out that maximal localisation states, $\ket{\psi_\xi^{\text{ML}}}$,  (i.e., those with the minimal uncertainty $\bra{\psi_\xi^{\text{ML}}} \Delta x \ket{\psi_\xi^{\text{ML}}}=\Delta x_0 = \hbar \sqrt{\beta}$) are and so these can be used to define what Kempf {\it et. al.} \cite{Kempf1995} call a quasiposition representation. They go on to show that the normalised maximal localisation states are, 
\begin{equation}\label{MLstate}
\psi_\xi^{\text{ML}} (p) = \sqrt{\frac{2\sqrt{\beta}}{\pi}} (1+\beta p^2)^{-\frac{1}{2}} \exp\left(-i\frac{\xi \arctan(\sqrt{\beta}p)}{\hbar\sqrt{\beta}}\right).
\end{equation}
We can then find the quasiposition wavefunctions, $\phi(\xi)$, by making use of scalar product (\ref{scalarproduct}),
\[
\phi(\xi)= \bra{\psi_\xi^{\text{ML}}}\phi \rangle.
\]
Interestingly, this algebra can also be seen to imply a modified dispersion relation. To show this we note that from eq. (\ref{scalarproduct}), position and momentum wavefunctions are still related by a (generalised) Fourier transform, and so we expect plane wave solutions when the momentum wavefunctions take on the form of a delta function. Using eqs. (\ref{scalarproduct}) and (\ref{MLstate}), the quasiposition wavefunction of a momentum eigenstate $\psi_{\tilde{p}}(p)=\delta(p-\tilde{p})$ with energy $E=\tilde{p}^2/2m$ is found to be,
\[
\psi_{\tilde{p}}(\xi)= \sqrt{\frac{2\sqrt{\beta}}{\pi}} (1+\beta \tilde{p}^2)^{-\frac{3}{2}} \exp\left(\frac{i\xi \arctan(\sqrt{2m\beta E})}{\hbar\sqrt{\beta}}\right),
\]
and so now the usual $2\pi/\lambda$ term in the exponential gains an additional functional dependence on the energy, 
\[
\lambda(E)=\frac{2\pi \hbar\sqrt{\beta} }{\arctan(\sqrt{2m\beta E})}.
\]

Unfortunately it turns out that the resulting dynamics from eq. (\ref{deformedcommutator}) and its generalisations is typically quite complex, and exact results are only known for a few cases. In particular, variations on the harmonic oscillator \cite{Kempf1995,Kempf1997,Chang2002,Dadic2003,Quesne2003,Quesne2004,Quesne2005} and for the 1-D coulomb potential \cite{Brau1999,Benczik2005,Stetsko2006,Stetsko2006a,Stetsko2008}. However, this type of analysis has still enabled attempts to bound $\beta$ by applying the resulting dynamics to macroscopic harmonic oscillators \cite{Bawaj2015}.

\subsection{Wheeler's quantum foam and decoherence}
\label{sec:Quantumfoam}

Wheeler interpreted relations such as (\ref{Wheeleruncertainty}),(\ref{Karolyhazyuncertainty}) and (\ref{KarolyhazyuncertaintyT}) as arising from fundamental fluctuations of space-time. On short scales, fluctuations in the metric could vary hugely, leading to the spontaneous creation of short lived particles or even wormholes \cite{Wheeler1962}. He famously compared the situation to an ocean, which when viewed from afar looks flat, but on closer inspection becomes choppy with foam forming on the smallest scales. These fluctuations, while typically much larger than the Planck scale, are still invariably tiny. For instance, the relation first put forward by K{\'a}rolyh{\'a}zy, (\ref{Karolyhazyuncertainty}), suggests that the uncertainty in measuring distances comparable to the observable universe ($\sim 10^{27}$m) is around $\Delta l \sim 10^{-15}$m -- far smaller than the size of an atom. Nevertheless, this still represents a significant potential advantage over the Planck length. In particular, it has been suggested that these space-time fluctuations lead to an intrinsic source of noise and can therefore potentially play a role the objective reduction of the state vector.

Interestingly, the relation (\ref{Karolyhazyuncertainty}) is not the only length uncertainty to appear in the literature (and indeed, as we noted earlier, there are also multiple suggestions for $\Delta g$ and $\Delta \Gamma$, and relation (\ref{Karolyhazyuncertainty}) itself implies $\Delta g \gtrsim (l_p/l)^{2/3}$ \cite{Ng1994}). Taking a phenomenological slant, these are sometimes parametrised as $\Delta l \gtrsim l_p^{\alpha} l^{1-\alpha}$ \cite{Amelino-Camelia1999,Amelino-Camelia2000,Amelino-Camelia2000a,Ng2005,Christiansen2006,Ng2008,Vasileiou2015}, where now the K{\'a}rolyh{\'a}zy/Holographic relation is represented by $\alpha=2/3$. Other popular models include $\alpha=1/2$ which is analogous to fluctuations induced by a random walk, while in the limit $\alpha=1$ we recover again the Planck length as the minimal uncertainty. In 1999, Amelino-Camelia \cite{Amelino-Camelia1999} suggested that some values of $\alpha$ could already be ruled out by gravitational wave interferometers. The general idea is as follows. While experiments like LIGO are sensitive to displacements smaller than $10^{-18}$m, this is still much too large to test space-time models directly. However, the displacement sensitivities of gravitational wave detectors are frequency dependent, where the noise is characterised by a spectral density $S(f)$ for frequency $f$. For a frequency band limited below by the inverse of the observation time, $t_{obs}$, the displacement noise is given by \cite{Amelino-Camelia1999,Ng2000},
\begin{equation}
\sigma^2 = \int^{f_{\max}}_{1/t} [S(f)]^{2} df.
\end{equation}
Assuming this noise arises from space-time fluctuations from the phenomenological model $\sigma = c\Delta t  \sim (ct)^{1-\alpha}l_{p}^{\alpha}$ then we expect the amplitude spectral density to be characterised by, 
\begin{equation}
S(f) \sim f^{\alpha - 3/2}c^{1-\alpha}l_p^{\alpha}.
\end{equation}
Data from the Caltech 40-meter interferometer (with a noise level of $3\times 10^{-19}$mHz$^{-1/2}$ near $450$Hz \cite{Amelino-Camelia1999,Ng2000}), for example, then appears to rule out the random walk model ($\alpha=1/2$).

As an aside, we can yet again note that fluctuations in space-time can also be viewed as energy-momentum uncertainties. Eq. (\ref{Karolyhazyuncertainty}), together with the de Broglie relation, $\lambda =h/p$ and the usual propagation of errors formula implies \cite{Ng2005},
\begin{equation}
\Delta p =\zeta p \left(\frac{p}{m_p c}\right)^{2/3}, \qquad \text{and} \qquad \Delta p =\gamma p \left(\frac{E}{m_p c^2}\right)^{2/3},
\end{equation}
where $\zeta$ and $\gamma$ are taken to be independent and $\sim 1$. If we assume the energy dispersion relation holds up to uncertainties in $p$ and $E$, then in the high energy limit $E\gg mc^2$ we find,
\begin{equation}
E^2 = m^2 c^4 + p^2 c^2 +2(\zeta-\gamma)\left(\frac{p}{m_p c}\right)^{2/3}.
\end{equation}

To understand how such fluctuations are actually supposed to lead to decoherence, it is worth examining a few of the notable models. In particular, these were motivated by the desire to address an equally troubling conceptual issue. Namely, the measurement problem.
\subsubsection{K{\'a}rolyh{\'a}zy model for gravitational decoherence}
The earliest decoherence model was proposed by K{\'a}rolyh{\'a}zy in 1966 \cite{Karolyhazy1966}.
As a space-time interval, $s$, in general depends on the metric $g_{\mu \nu}$ through $ds^2= g_{\mu\nu} x^{\mu}x^{\nu}$, K{\'a}rolyh{\'a}zy viewed relation (\ref{Karolyhazyuncertainty}) (in the limit $s=cT$) as arising from a smearing of space-time. He proposed that we should instead consider a family of matter-free metrics, $g_{\mu\nu}^{\beta}$, distributed in such a way that when space-time intervals are appropriately averaged over $\beta$, we recover both the line element $s$ and the uncertainty $\Delta s$,
\[
s= \avg{s_{\beta}}, \qquad\qquad \Delta s = \avg{(s-s_{\beta})^2}^{1/2}.
\]
In lieu of a quantised metric, here $\avg{}$ now refers to a stochastic average. Given these conditions, and that the minimum resolution is expected to be small, it is natural to describe the smeared space-times as fluctuations around the Minkoswki metric. For $v\ll c$ we need only care about the component,
\[
(g_{00})_{\beta}=1+\gamma_{\beta}(x,t).
\]
The matter-free condition then implies $\gamma_{\beta} (x,t)=0$ and so the fluctuations are assumed to take the form of solutions of a wave equation. By assuming that the Fourier coefficients are independent stochastic variables, which average to zero as $\beta$ varies, K{\'a}rolyh{\'a}zy was able to find explicit solutions satisfying eq. (\ref{Karolyhazyuncertainty}) \cite{Karolyhazy1966,Karolyhazy1986}. 

The family of metrics imply that for a spinless particle, a more appropriate description of the dynamics is through a generalised Klein Gordon equation. However, for each $\gamma_{\beta}\neq 0$, the non-relativistic limit now no longer recovers the usual Schr\"odinger equation, but instead gains an additional potential of the form $V_{\beta}(x,t)=Mc^2\gamma_{\beta}(x,t)$\footnote{Where the fluctuations are assumed small, and we also must factor out a $e^{-(i/\hbar)Mc^2}$ from the wavefunction.}. A wavefunction propagating through this smeared space-time is then assumed to pick up a phase difference with respect to one in the Minkowski metric. The variance of the relative phase between points (now calculated using the explicit solutions for $\gamma_{\beta}$) give a measure of decoherence. K{\'a}rolyh{\'a}zy proposed that when this phase is on the order $\pi$ the wavefunction loses coherence and undergoes a stochastic reduction. The corresponding spatial separation is then referred to as the coherence length, $a_c$, with the timescale for a wavefunction to grow to this extent given by the decoherence time $\tau_c$. For a single particle these are given by,
\begin{equation}\label{Kcoherencelength}
a_c\approx L_c\left(\frac{L_c}{l_p}\right)^2, \qquad \qquad \tau_c \approx \frac{Ma_c^2}{\hbar}.
\end{equation}
Where $L_c$ is the Compton wavelength, $L_c=\hbar/Mc$. To give an example, an electron would then undergo stochastic reduction once its wavelength exceeded $a_c\approx 1.8 \times 10^{33}$m, or equivalently after a time $\tau_c \approx 10^{70}$s, which far exceeds the age of the Universe.

However, this approach for calculating the metric fluctuations has been criticised, on the grounds that the Fourier expansion of $\gamma_{\beta}$ should require a cut-off in $k$-space to avoid divergences. K{\'a}rolyh{\'a}zy himself had recognised this, and suggested a cut-off should be on the order of $k_{\max}=10^{13}\text{cm}^{-1}$. However Diosi and Lukacs argued that this leads to extremely unphysical cosmological energy densities, which would be many orders of magnitude higher than that of a neutron star \cite{Diosi1993} (though see also the reply by \cite{Ng1997}). A slightly different approach, claiming to sidestep these issues, but still recovering the same results as \cite{Karolyhazy1966} was presented by Frenkel \cite{Frenkel2002}. He instead modeled the clock uncertainty as a fluctuation in measured time\footnote{He considered time as an as yet unknown local operator which he surmised would emerge in a future theory of quantum gravity. Averages, and variances (including those appearing in relation (\ref{Karolyhazyuncertainty}) should then be calculated as quantum mechanical expectation values on this theory's vacuum state. It turns out, that as long as this is applied consistently throughout the K-model, the precise details do not matter. This allows the coherence length to be calculated directly from other variances.},
\begin{equation}
t(x,t)= t+\tau(x,t),
\end{equation}
Which should be consistent with the time interval version of eq. (\ref{Karolyhazyuncertainty}),
\begin{equation}\label{KarolyhazyuncertaintyT}
(\Delta T)^3 \gtrsim t_p^2 T.
\end{equation}
As before, the model assumes that a measurement of $t(x,t)$ returns the expectation value, which should be the familiar time $t$. This immediately implies that $\avg{\tau(x,t)}=0$ and that $\avg{T(x,t)}=t_2-t_1$. It is clear the fluctuations in time can be formally written as a function of the metric fluctuations, and so it is natural to assume a propagating wavefunction should pick up a similar relative phase. However, rather than invoking a curved space-time explicitly, Frenkel argued that the required phase effectively emerges from the (usually position independent) rest mass energy contribution to the Hamiltonian. Recall, that for an isolated pure state, the wavefunction evolves as, 
\begin{equation}\label{Schrodingertimeevolution}
\psi(x,t)=e^{-\frac{i}{\hbar}Ht} \psi(0,t).
\end{equation}
In normal, non relativistic quantum mechanics, the rest energy (for a single particle) amounts to a coordinate independent phase $\Phi(t)=(i/\hbar)mc^2t$ which can safely be neglected in the above. However, in the K{\'a}rolyh{\'a}zy model, a wavefunction displaced in time now picks up an additional space-time dependent phase and so $\Phi(t)\rightarrow\Phi(x,t)$. Strictly speaking, one should also worry about whether we should also use $t(x,t)$ in the Schr{\"o}dinger equation and the wavefunction, but in terms of phase contributions, the rest mass energy should be much larger than that from the kinematic contributions. Thus, as an approximation, to any solution of the Schr{\"o}dinger equation the K{\'a}rolyh{\'a}zy model associates a state, 
\[
\psi_K(x,t) = e^{i\Phi(x,t)} \psi_0(x,t),
\]
To calculate the variance $\Delta^2_{\Phi}(x,x',t)=\langle ( \Phi_R(x,x',t)-\langle\Phi_R(x,x',t)\rangle)^2\rangle$, we note that the condition $\avg{\tau(x,t)}=0$ implies that the expectation $\avg{\Phi(x',t)-\Phi(x,t)}$ vanishes. Therefore for a single particle, 
\begin{equation}
\Delta^2_{\Phi}(x,x',t)=\frac{c^4 M^2}{\hbar^2} \avg{[\tau (x',t)-\tau (x,t)]^2}.
\end{equation}
The same condition also implies the right hand side is equal to the variance in the time difference, $T_{\text{sync}}=t(x',t)-t(x,t)$, between two synchronised clocks separated by a distance $a=|x'-x|$\footnote{Again, we are assuming $v\ll 1$.}. This can be estimated from relation (\ref{KarolyhazyuncertaintyT}) by considering a synchronisation procedure which involves a light signal sent from one clock to the other, and back again. The uncertainty in the return time $2a/c$ will be subject to the same uncertainty (\ref{KarolyhazyuncertaintyT}). Thus, 
\[
(\Delta T_{\text{sync}})^2 \approx (\Delta T)^2 \gtrsim \frac{l_p^{4/3}}{c^2}a^{2/3},
\]
and so finally the spread in relative phase is found to be, 
\begin{equation}
\Delta_{\Phi}(a)=\frac{l_p^{2/3}}{L} a^{1/3}.
\end{equation}
Setting $\Delta_{\Phi}(a)\approx \pi$ recovers the coherence length (\ref{Kcoherencelength}). The decoherence time is then estimated by considering the free expansion of a minimum uncertainty state, $t_c \approx m a_c^2/\hbar$. Extending to multi-particle states is straightforward \cite{Frenkel2002}, and with suitable approximations one can calculate the corresponding quantities for homogeneous spherical objects of radius $R$. These turn out to be, up to some numerical factors,
\begin{equation}
a_c \approx \left\{
\begin{aligned}
& \left( \frac{R}{l_p} \right)^{2/3} L, \qquad \qquad a_c \ll R \\
& \left( \frac{L}{l_p} \right)^{2/3} L, \qquad \qquad a_c \gg R
\end{aligned}
\right.
\end{equation}
In the transition region, where $a_c \approx R$, we can estimate $R\approx \hbar^2/Gm^3$. For a ball of density $1gcm^{-3}$ this suggests the transition sizes and masses marking the point between microscopic and macroscopic behaviour are $a_c^{\text{tr}} \approx 10^{-5}$cm and $M^{\text{tr}} \approx 10^{-14}$g.

\subsection{Di\'{o}si and the ``Schr\"{o}dinger-Newton" equation}
\label{sec:diosi}
A second model, now relying directly on fluctuations of the gravitational field, was put forward by Di\'{o}si \cite{Diosi1984,Diosi1987}. He first made an argument for using a nonlinear modification of Schr\"{o}dinger equation to incorporate gravity (what Penrose would later call the Schr\"{o}dinger-Newton equation (SN)). His line of reasoning was roughly as follows:
In quantum mechanics, the wavepackets associated to free particle solutions tend to spread out in time, in contrast to the kind of localisation seen in classical counterparts. He then argues that while the wavefunctions corresponding to macroscopic masses, which are initially localised within a large enough volume (say on the order of that occupied by an atom) may not spread out appreciably over large timescales, the centre of mass of collections of atoms can be defined more accurately. Thus, macroscopic states would quickly exhibit non stationary behaviour. Instead one could look for a modification to the usual Schr\"{o}dinger equation, which could act to localise states directly. Di\'{o}si proposes that the source of this modification is gravity, and should be incorporated into the linear Schr\"{o}dinger equation by the addition of a Newtonian gravitational term\footnote{As an aside, the SN equation as argued by Di\'{o}si, is derived from the semi-classical Einstein equations, where in the Newtonian limit the source is given by the expectation value of the mass density. The Schr\"{o}dinger-Newton equation can then be taken to represent a bona fide description of a single particle. On the other hand, one can instead assume that gravity is also quantised, and so that the fundamental Schr\"{o}dinger equation remains linear. The Schr\"{o}dinger-Newton equation can then be derived as an approximation to gravitational interaction for systems with large numbers of particles. See, for example, the paper by Bahrami et al. (2014) for an overview.},
\begin{equation}\label{SN}
i\hbar \frac{\partial \Psi}{\partial t}= -\frac{\hbar^2}{2m}\nabla^2 \Psi + m\phi\Psi,
\end{equation}
where the potential is a solution to the Poisson equation,
\[
\nabla^2 \phi=4\pi mG|\Psi|^2.
\]
While not exactly solvable, this equation has been studied both qualitatively and numerically in the literature. In particular, the modified equation preserves normalisation, momentum and energy, and solutions are equivalent (up to a phase) under Galilean transformations \cite{Moroz1998}. Di\'{o}si then estimated that the ground state of a Gaussian wavefunction will have a characteristic width on the order of,
\[
a_0\approx\frac{\hbar^2}{Gm^3},
\]
Which coincided with the demarcation between macroscopic and microscopic scale given by K{\'a}rolyh{\'a}zy.

To develop this model further, Di\'{o}si and Lukacs \cite{Diosi1987a} then argued that fluctuations in space-time mean that the gravitational acceleration $g$ cannot be measured to arbitrary accuracy. Instead, they find the uncertainty in the average measurement (in time and space) of the field $\bar{g}$ in time $T$ using a device of volume $V$ is roughly, 
\[
(\delta \bar{g})^2\sim  \frac{\hbar G}{VT}.
\]
Thus the potential term appearing in the Schr\"{o}dinger-Newton equation above should now be stochastic, with the gravitational field now possessing fluctuations with a spread of the order of the measurement indeterminacy \cite{Diosi1987}. The state itself then becomes a stochastic variable, with its corresponding density function obeying a master equation. This turns out to have a characteristic damping,
\begin{equation}\label{Diosit}
\tau_d^{-1} = \frac{G}{\hbar}\int dr dr'\frac{[\rho(r|X)-\rho(r|X')][\rho(r'|X)-\rho(r'|X')]}{|r-r'|}, 
\end{equation}
where $X$ denotes the configuration coordinates of a system, and $\rho(r|X)$ is the mass density at a point $r$. The natural analogue of K{\'a}rolyh{\'a}zy's $a_0$ is the critical coherent width $l_{crit}$ which for a rigid sphere is now a function of $R$. However, when $l_{crit}=R$ it turns out the value still coincides with the coherence length, $a_0$ given above.

\subsection{Intrinsic decoherence}
\label{sec:milburn}
A more model independent approach is to seek a modification to Schr\"{o}dinger's equation such that, as the system approaches a macroscopic level, coherence is destroyed. An example of this type comes from Milburn \cite{Milburn1991} who postulates that evolution of a system for sufficiently short timesteps is not unitary; instead, it is a stochastic sequence of identical unitary transformations. This adopts the idea of the universe having a minimum time scale. Using the frequency of these unitary steps, labelled $\gamma$, as an expansion parameter, it is possible to see that with large enough frequency the evolution appears continuous at laboratory time scales. Standard quantum mechanics is recovered at zeroth order of the expansion while decoherence in the energy eigenstate basis is witnessed at first order.

In more detail, \cite{Milburn1991} takes the standard quantum mechanics formulation of the evolution of a state,
\begin{equation}
\label{eq:MilburnEvolution}
\rho(t+\tau) = e^{-i{H}\theta(\tau)/\hbar}\rho(t) e^{i{H}\theta(\tau)/\hbar},
\end{equation}
and replaces it with three postulates:
\begin{enumerate}
\item The change of the state of a system is uncertain on a sufficiently small time scale, with the probability of change given by $p(t)$.
\item The change of the state of a system is described by (\ref{eq:MilburnEvolution}). In ordinary quantum mechanics, $p(\tau)=1$ and $\theta(\tau)=\tau$. In this generalisation, it is only the case that $p(\tau)\rightarrow 1$ and $\theta(\tau)\rightarrow \tau$ for $\tau$ large.
\item There is a minimum unitary phase change, implied by $\lim_{\tau\rightarrow 0}\theta(\tau)=\theta_0$.
\end{enumerate}

In particular, this approach introduces an intrinsic decoherence to a system which may then be associated to gravity through the choice of parameters. The model can equally be generalised to stochastic space instead of time \cite{Milburn2006}.
\\
\\
It should be noted, however, that an argument for gravity's role in the state reduction can come from a different perspective. Such a ``model" was proposed by Penrose \cite{Penrose1986,Penrose1996,Penrose1998,Penrose2000,Penrose2002,Penrose2004,Penrose2014}, and shall be discussed in general terms below.

\subsubsection{The Penrose model for gravitational decoherence}
\label{sec:penrose}

In contrast to those introduced in the previous sections, the motivation does not explicitly reference fluctuations of space-time. Instead the argument lies on an apparent fundamental incompatibility between the principle of superposition on the one hand, and general covariance on the other. In general relativity, space-time has ``no prior geometry", that is we can only attribute physical meaning to points on a manifold after the metric has been dynamically determined -- these coordinates do not exist independently (this is the essence of Einstein's famous ``hole problem" \cite{Stachel1993,Christian2001}). It follows from this that there is no way for a pointwise identification of two space-times. The difficulty then arises when one tries to form a superposition of two masses at "different" locations. Quantum theory implicitly assumes an a priori background space-time with a consistent notion of time translation. However, when we attempt to incorporate the diffeomorphism invariance of general relativity this is no longer true. Penrose claims this ambiguity in the identification of time evolution, when space-times are "significantly different", manifests as an energy uncertainty, which in turn leads to a decay of the superposition much in the same way as a radioactive nuclei.

Clearly, without a full theory of quantum gravity, it's not obvious how one should account for the misidentification. However, one can make an approximation by considering flat space-times in the limit of Newtonian gravity. A variety of criteria have been put forward for how we should measure the difference. One of the earliest was to make use of an integral by Fierz, evaluated over a timelike hypersurface "between" two lumps of matter. This turns out to be something of the order of a time multiplied by the gravitational energy gained by moving one lump from coincidence to separation, while in the field of the other. Differing spacetimes are those for which this integral is of order $1$, and so a natural timescale for this to occur is given by the inverse of the associated energy. Later, Penrose \cite{Penrose1996} rephrased this idea by taking the (square of) the difference between the gravitational force per unit test mass felt at a point, integrated over all space, as a measure of incompatibility of the two spacetimes. With a suitable use of constants, he turns this into an energy and invokes a Heisenberg-like uncertainty relation to set a timescale for the reduction effect,
\begin{equation}
\tau_d^{-1} \approx \frac{E_G}{\hbar} = \frac{4\pi G}{\hbar}\int dx^3 dy^3\frac{[\rho(x)-\rho'(x)][\rho(y)-\rho'(y)]}{|x-y|}, 
\end{equation}
where $E_G$ is the gravitational self energy of the difference between the the mass distributions of two superposed stationary states. This bears a remarkable similarity to the timescale suggested by Di\'{o}si, eq. (\ref{Diosit}). To be clear, there are a couple of points we should take care to emphasize. First, the states corresponding to the lumps from which we construct the superposition should be 1) stationary in their own right, and 2) of equal energy. If not, then the superposition need not be stationary (c.f. Rabi oscillations). Secondly, as what we are concerned with is an error in the identification of the time evolutions (more strictly, the Killing vectors) associated with the two space-times, one may take this as an uncertainty in the energy ($i\hbar d/dt$).

A more constructive argument makes explicit use of the weak equivalence principle to connect solutions of the Schr\"odinger equation in a free fall frame to a lab frame in constant gravitational field. This leads to an overall phase difference proportional to $t^3$ \cite{Penrose2009,Penrose2014}. We then consider the effect on a unit test mass in the vicinity of the superposed lumps and integrating over all space we effectively recover the error measure in terms of the gravitational self energy postulated earlier (again with suitable addition of multiplicative constants). 

One issue with this criteria is that point-like masses lead to a divergence in the expression for $E_G$ when evaluated at that point. A possible solution to this would be to introduce a finite resolution, as suggested by Di\'osi. Rather than taking this approach (and therefore introducing an additional fundamental length scale), Penrose considered another option. Since superposed states must be stationary wavefunctions, a single point mass must have as a stationary solution a plane wave with no localised position. As a result, it is not clear what the gravitational self energy should be. To rectify this, we can consider two cases. First, there could be some self attraction between the point mass and its own wavefunction. Second, we consider replacing the point mass with a lump composed of multiple particles weakly binding the state to the centre of mass. In both of these cases, Penrose proposes that the relevant description is provided by the Schr\"odinger-Newton equation (\ref{SN}), whose stationary solutions give the possible final reduction states. 

It is worth discussing a little why the SN equation might be relevant. Initially considered by \cite{Ruffini1969} in the context of self gravitating particles in a boson star, Penrose used much the same approach in the treatment of displaced lumps of matter, although in his case the smaller scale made taking the Newtonian limit less ambiguous. Penrose used the natural stationary states of the SN equation to estimate how different two space-times are, and notes that these should also be the states for which the wavefunction collapses. While this is true, it should not change how we estimate the difference in space-times. To estimate this difference we also need to find the gravitational self energy, $E_G$, of the difference in mass distributions of these solutions. A true timescale for this collapse is unknown at this point, as the current approach instead considers sufficiently large objects such that the classical mass distribution is a reasonable approximation. See however the discussion by \cite{Adler2007}.

There are, however, some obvious criticisms. To begin, it is argued that since the SN equation is non-linear, it should violate special relativity \cite{Gisin1989}. From one perspective, this does not rule out Penrose's approach as the equation seems necessary only to provide stationary states for lumps of mass. An alternative description may solve this problem, although in either case it is desirable to have some quantum mechanical description of matter (in its own space-time) accounting for our intuition that the wavefunction should be in some sense localised. There still remains an implicit indeterminacy to which state the wavefunction collapses, and so a full theory incorporating Penrose's ideas may be tenable. Secondly, Penrose himself mentions that the considerations above do not tell us anything about how to treat the wavefunction of a single isolated particle as it spreads through space. That is, it does not deal with the question of whether or not a spontaneous reduction should take place (though the SN equation has been put forward by Di\"osi for precisely that purpose). On the other hand, \cite{Christian2001} argues that the appropriate setting to consider in the non-relativistic limit is the Newton-Cartan formalism. Here the SN equation falls out as a variation of the Lagrangian density. However, he also argues that this setup requires an additional condition, namely that there is an absolute rotation (i.e. there is no rotation between inertial frames). In this case, the SN is invariant under transformations, up to a phase, and so he expects to see no Penrose-collapse in the types of experiment suggested so far. On the other hand, Penrose claims the explicit form of this phase implies an inequivalence of the vacua of the two solutions, thereby recovering his standard expression for $E_G$.

\section{Experimental proposals}

Since Amelino-Camelia's suggestion that gravity-wave detectors and gamma ray bursts could place bounds on the energy-momentum dispersion relations \cite{Amelino-Camelia1999,Amelino-Camelia1998} (see also section \ref{sec:Quantumfoam}), there have been a number of proposals for testing the effects discussed so far. A large proportion of these new tests are based in field of experimental quantum optics. Unsurprisingly, modified dispersion relations have received attention here also, in particular via the measurement of the recoil frequency of cold atoms involved in two photon Raman transitions \cite{Amelino-Camelia2009,Mercati2010}. In contrast to astrophysical observations, these low energy experiments allow to test the non-relativistic limit ($p\ll mc$), characterised by relations of the form\footnote{Equally, the data could be also applied to models assuming a different form.}, 
\begin{equation}
E \approx m c^2 + \frac{p^2}{2m} + \frac{1}{2m_p} \left( \xi_1 mcp + \xi_2 p^2 + \xi_3 \frac{p^3}{mc} \right).
\end{equation}
Notably, existing data on cesium experiments have already placed bounds on the $\xi_1$ parameter, where \cite{Amelino-Camelia2009} reports ``$-6.0<\xi_1<2.4$ at the $95\%$ confidence level". Of course, this also includes the possibility that $\xi_1=0$. Tests of a similar parameterisation using interferometry with nanodiamonds have also been put forward \cite{Albrecht2014}.

Bekenstein has recently proposed another tabletop experiment, this time using optomechanics to test Wheeler's notion of quantum foam \cite{Bekenstein2012, Bekenstein2014}. When a single blue photon of wavelength $445$nm hits a dielectric slab of mass $0.15$g, the slab is moved by a distance on the order of the Planck length. Assuming a refractive index of $1.6$, we can calculate that $89.6$\% of the photons are transmitted. However when the shift of the centre of mass is around $l_p$, then from Wheeler's arguments we expect that space-time is no longer smooth and the translation of the slab should be impeded.  The impact will be a lowering of the number of photons measured at the egression side of the slab. While Bekenstein was not able to estimate the change of the transmission rate he expected the effect to be small.

In fact, as we pointed out earlier, optomechanics has also formed the basis of a number of other proposals (and indeed has recently been suggested as route for testing the Schr\"odinger-Newton equation \cite{Grosardt2015}). Here we discuss two in detail.   

\subsection{Optomechanical tests for decoherence of the wavefunction}

\begin{figure}
\centering
 \includegraphics[width=9cm]{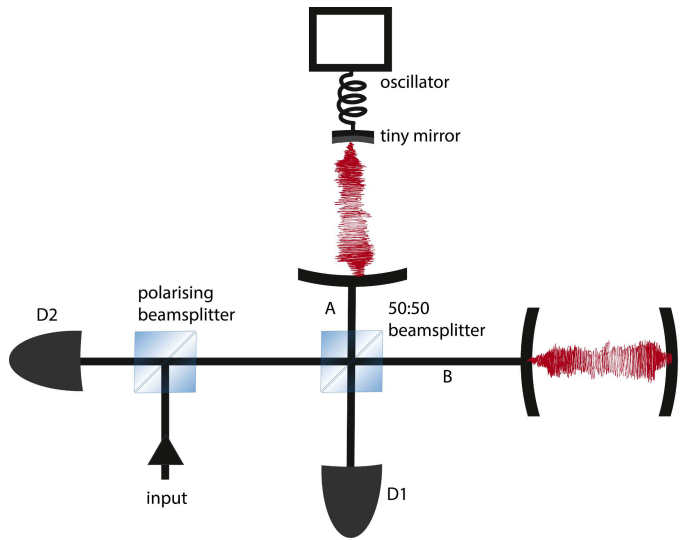}
    \caption{Experimental set-up of the proposal in \cite{Marshall2003}}
  \label{fig:MarshallExp}
\end{figure}

The first concerns a notable proposal put forth by Marshall {\it et. al.} for testing the decoherence of a macroscopic superposition  \cite{Marshall2003}. This proposal can be though of as a refinement to some of Penrose's original suggestions \cite{Penrose2002} (which themselves were derived from discussions with Johannes Dapprich, Anton Zeilinger and Anders Hansson \cite{Penrose2004}). The setup makes use of work by \cite{Mancini1997}, and \cite{Bose1997,Bose1999}, among others, and is focused on producing a macroscopic superposition out of the position states of a mirror using the radiation pressure from a field. This is achieved as follows (see Fig. \ref{fig:MarshallExp}).

A mirror is attached to a cantilever in a cavity of length $L$ . A photon incident on the mirror will impart some momentum on the surface, thereby increasing the length of the cavity to $L+x$. This decreases the resonant frequency, $\omega_c=\pi nc/L$, to $\omega_c'=\pi n c/(L+x)$. Thus in the limit of small displacements, the Hamiltonian, $H=\hbar \omega_c a^{\dagger}a$, picks up a factor $\approx 1-x/L$, where x is the displacement of the mirror. The total Hamiltonian for the system (in the weak coupling limit) is then given by \cite{Mancini1997},
\begin{equation}\label{HamiltonianMirror}
H=\hbar \omega_c a^{\dagger}a + \hbar \omega_m b^{\dagger}b -\hbar G \hbar a^{\dagger}a (b+b^{\dagger}),
\end{equation}
where the coupling constant $G=(\omega_c/L)\sqrt{\hbar/2M \omega_m}$, and $\omega_m$, $M$ and $b$ are the frequency, mass and phonon creation operator for the mirror. If the field in the cavity is initially in a Fock state $\ket{n}_c$, then the ground state of the mirror will evolve as,
\[
\ket{0}_m \rightarrow \ket{\kappa n(1-e^{-i\omega_m t})}_m,
\]
where $\kappa=G/\omega_m$. To generate the desired macroscopic superposition, we need only prepare the initial cavity field as a superposition of Fock states. The earlier work by \cite{Bose1997,Bose1999} did not address a particular method for this preparation, however Marshall {\it et. al.} \cite{Marshall2003} proposed the use of two cavities in each arm of a Michelson interferometer. The oscillating mirror is located in one arm, which we will label $A$. The initial state is then $\psi(0)=(1/\sqrt{2})(\ket{0}_A\ket{1}_B+\ket{1}_A\ket{0}_B)\ket{0}_M$, which after time $t$ evolves to\footnote{The additional phase factors come about from the time evolution of the cavity field in arm $A$. This can be seen from the time evolution operator corresponding to the Hamiltonian (\ref{HamiltonianMirror}) is (\cite{Mancini1997}, \cite{Bose1997}),
\[
U(t)=e^{-i\omega_c a^{\dagger}at}  e^{i\kappa^2(a^{\dagger}a)^2(\omega_m t-\sin \omega_m t)}  e^{\kappa a^{\dagger}a(\eta b^{\dagger}-\eta^* b)}  e^{-ib^{\dagger}b \omega_m t},
\]
where $\eta=1-e^{-i\omega_m t}$.}  
\[
\ket{\psi(t)}=\frac{1}{\sqrt{2}} e^{-i\omega_c t} [\ket{0}_A\ket{1}_B\ket{0}_m + e^{i\kappa^2(\omega_m t-\sin \omega_m t)} \ket{1}_A\ket{0}_B \ket{\kappa (1-e^{-i\omega_m t})}_m ]. 
\]
There are a few important points to mention here. First, if the photon is in arm $A$ then the mirror oscillates with a displacement proportional to the coupling strength, $\delta x(t)=2\kappa (1-\cos\omega_mt)\sqrt{\hbar/2M\omega_m}$. We then see that the initial superposition state of the cavity field sets up a superposition of spatially separated mirror states (which clearly must be regarded as macroscopic). In the absence of decoherence, the mirror returns to its original position after a full period, $t=2\pi/\omega_m$. There are different approaches to determining the decoherence of this system. \cite{Bose1997} showed that by incorporating photon loss in the cavity, along with a generic, model-independent loss of coherence between the spatially separated mirror superposition (assumed to arise through environmental induced decoherence on the mirrors motion), one can measure the decoherence rate, $\Gamma_m$, of the mirror by determining the probability of the cavity field to be in the state $\ket{+}=(1/\sqrt{2})(\ket{0}+\ket{1})$ at $t=2\pi/\omega_m$. In particular, this can be accomplished by mapping the state of the field onto a two level atom passed through the cavity such that its time of flight is half a Rabi oscillation period. Furthermore, it can also be shown that the state of the cavity field at time $t=2\pi/\omega_m$ is independent of the precise initial state of the mirror (so long as it is in a thermal state), and so one can still study the decoherence of macroscopic superpositions even if these states are not pure Schr\"odinger cat states. 

On the other hand, a more direct measurement of decoherence can accomplished using the setup of \cite{Marshall2003}. Here we can rely on the interferometer itself to observe interference of the photon once it leaves either cavity. This is given as function of the off-diagonal terms of the photon's reduced density matrix, which are again susceptible to the generic decoherence mechanisms outlined above. 

In practice, the mirror will initially be at some finite temperature, and so the maximum interference visibility will be very tightly distributed around $t=2\pi/\omega_m$, and zero otherwise\footnote{The width of the peak is narrowed by a factor $\sqrt{\bar{n}}$. \cite{Marshall2003} propose a cooling scheme to bring the mirror closer to its ground state (and below the temperature of the environment $T_E$).}. Therefore, in both cases, the rate of environmentally induced decoherence (EID) must be at most on the order of $\omega_m$ otherwise the mirror superposition will have decohered completely before a measurement of $\Gamma_m$ can take place. This is the central condition which {\em must} be satisfied in order for these experiments to be feasible. The simplest (and best understood) estimate is to assume an ohmic environment of harmonic oscillators. The decoherence rate for off-diagonal terms is then roughly (e.g. \cite{Caldeira1983}),
\[
\Gamma_m=\frac{1}{\hbar^2}\gamma_m k_BT_E M (\delta x)^2,
\]
where $\gamma_m$ is the damping rate for the mirror and $T_E$ is the temperature of the environment. One should be careful with the conditions of validity of this equation, however it is to be taken as an order of magnitude approximation. 

Furthermore, in anticipation of testing gravitational decoherence, one should expect the states of the mirror are to some extent macroscopically discernible\footnote{Strictly speaking, this is perhaps a weaker condition, but it does help to simplify the analysis.}. This is most easily captured by requiring the separation $\delta x(2\pi/\omega_m)$ of the mirror's coherent states to be greater than their width, i.e, $\delta x \sim \sqrt{\hbar/M\omega_m}$ or, $\kappa^2 \gtrsim 1$\footnote{Note, \cite{Bose1999} impose a further condition that $\delta x$ is greater than the thermal de Broglie wavelength, which is a requirement for the decoherence time expression to be valid.}. \cite{Marshall2003} apply this condition to the decoherence rate for an ohmic environment and then demand that this should be less than the mirror frequency. This implies that the quality factor for the mirror, $Q=\omega_m/\gamma_m$, must be roughly greater than, 
\[
Q\gtrsim \frac{k_B T_E}{\hbar \omega_m}.
\]
We also need to assess whether such a separation is feasible in the first place. In one full period, the photon will make $N=2\pi c/(2\omega_m L)$ round trips. Substituting for $\omega_m$ in the condition $\kappa^2 \gtrsim 1$, and recalling that $\omega_c=2 \pi c/\lambda$, leads to, 
\[
\frac{2\hbar N^3 L}{\pi c M \lambda}\gtrsim 1.
\] 
These are the two constraints necessary to observe decoherence of a macroscopic superposition under the ohmic environment model\footnote{A similar set of conditions is used by \cite{Bose1999}.}. By choosing to work in the optical regime, and assuming various values corresponding to the, then current, state of the art \cite{Bose1999} conclude that a test of EID of macroscopic superpositions can be realised if mirror reflectivity can be improved by approximately $3-5$ orders of magnitude. 

It is at this point that conclusions start to diverge. Bose \etal \cite {Bose1999} proceed to make an estimate for the Penrose decoherence rate, $\tau_{d}^{-1}\sim E_G/\hbar$. They base this on calculating the Newtonian gravitational self energy of two displaced (spherical) mirrors, where the separation $\delta x$ is less than the radius, $R$, of the mirror. To leading order, they take,
\[
E_G\sim \frac{GM^2(\delta x)^2}{R^3}.
\]
(c.f. with \cite{Penrose2014} page 571, for example). Combining with the expression for $\delta x$ above, and re-expressing $R^3\propto M/\rho$, (where $\rho$ is the density) they find that the gravitational decoherence rate will dominate the environmental decoherence rate when, 
\[
G\hbar\rho \gtrsim k_B T_E \gamma_m .
\]
Assuming typical solid densities, and optomistic values $T_E=0.1K$ and $\gamma_m =10^{-2} s^{-1}$ requires and improvement of $T_E\gamma_m$ by $16$ orders of magnitude. A similar conclusion is drawn by Adler \cite{Adler2007}, who instead calculates $E_G$ for a cube of side $S$ displaced by a small fraction of its dimensions. In the Newtonian limit, he obtains
\[
E_G=\frac{4\pi}{3} G (\delta x)^2 S^3 \rho^2 = \frac{4\pi}{3}\frac{ G m^2(\delta x)^2}{ S^3 }.
\]
For a $10\times10\times10\mu$m mirror of mass $5\times10^{-12}$kg, and a separation $~10^{-13}$m, this leads to a gravitationally induced decoherence rate of, 
\[
\tau_d^{-1}\approx 6.7 \times10^{-10} s^{-1}.
\]
This is much slower than the EID rate for the realistic parameters assumed by \cite{Marshall2003}, 
\[
\Gamma_m \approx 2.5\times 10^3 s^{-1},
\]
and still some $13$ orders of magnitude off. In contrast, Marshall \etal  \cite{Marshall2003} claim that the necessary $Q/T$ improvements needed to test Penrose idea is only $6$ orders of magnitude. One could argue that the Newtonian calculations of the gravitational self energy is incorrect, for the reasons outlined by \cite{Penrose1998,Penrose2000,Penrose2014}. Nevertheless, an evaluation of $E_G$ by means of the Schr\"odinger-Newton equation has not been carried out. Instead, they appear to make use of a approximate value given in \cite{Penrose2000},
\[
\tau_d^{-1}\approx \frac{\hbar}{20G\rho^2 R^5}.
\]
This expression ignores the actual displacement, and assumes that it should be at least of the order of the diameter of the objects (modelled as spheres). Even assuming $\delta x$ is on the order of the mirror dimensions, this still gives a decoherence rate of $\sim 3\times 10^{-8}s^{-1}$, and clearly with the increased separation, the environmental decoherence rate should drastically increase. It therefore seems difficult to understand the basis for the claim that an improvement in $Q/T$ of only 6 orders of magnitude is required. Nevertheless, over the last 10 years, these types of experiments have undergone significant refinements \cite{Pepper2012}, and so one can hope they will soon reach sensitivities required to test the ideas outlined in the previous section. 

However, one should make clear that gravitational decoherence falls within a much wider literature on collapse models, most of which could also be tested using macroscopic superposition experiments such as the one described above. A notable example would be the GRW model \cite{Ghirardi1986} (and the related CSL model). For a review of these ideas, see \cite{Bassi2013}. In principle, experimental discrimination of competing theories is difficult (particularly in light of problems evaluating $E_G$). However at least in the case of intrinsic vs. extrinsic (for example, arising from a quantised gravitational field) it has recently been suggested that one may be able to distinguish between them using dynamical decoupling \cite{Arenz2015}.

\subsection{Testing deformations of the commutation relations}
The second proposal concerns a test of a potential deformation of the canonical commutation relations. As pointed out in section \ref{deformedcommutators}, if the uncertainty principle has to be modified, the modification should be reflected in the commutation relation between the position and momentum operators because the lower bound of the uncertainty is determined by the commutator value \cite{Robertson1929}:
\begin{equation}
\Delta x\Delta p\geq {1\over 2}|[{x},{p}]|
\label{Robertson}
\end{equation}
For instance, if $\Delta x\Delta p\geq \hbar (1+\beta_0(\Delta p/(M_pc)^2)/2$ where $\beta_0$ is a parameter that quantifies the modification strength. Various theories in quantum gravity expect that $\beta_0$ is around 1, which makes the modification factor extremely small. The accuracy of the current-state-of-the-art experiments is to reach the value of $\beta_0$ around $10^{34}$ at facilities such as GEO 600 \cite{Grote2008}.
Pikovski {\it et al.} \cite{Pikovski2012} proposed to measure the modification value to an unprecedented accuracy based on optomechanical model similar to the setup described above to measure the decoherence. Instead of the continuous interaction, they consider the pulsed interaction between optical fields and a nanomechanical oscillator. In that case, the effective Hamiltonian is the interaction part of the Hamiltonian: 
\begin{equation}
{H}_I=-\hbar g{n}{x}
\label{int-Hamiltonian}
\end{equation}

From the Hamiltonian it is clear that due to the position of the oscillator, the phase of the optical field is shifted and the optical field pushes the oscillator by adding extra momentum to the oscillator. In quantum mechanics this interaction correlates the states of the oscillator and the field. Consider a series of pulsed interactions with free evolution of the oscillator for the duration of $T/4$ where $T$ is the period of the oscillator. Due to the fact that the free evolution transfers $x\rightarrow p \rightarrow -x \rightarrow -p$, the evolution of the total system is represented by
\begin{equation}
{U}= \mbox{e}^{-ig{n}{p}t}\mbox{e}^{-ig{n}{x}t}\mbox{e}^{ig{n}{p}t}\mbox{e}^{ig{n}{x}t}.
\label{evolution}
\end{equation}
As ${x}$ and ${p}$ are operators which do not commute, the operations do not cancel each other. According to the Baker-Hausdorf Theorem \cite{Barnett1997}, the total operation becomes ${U}=\exp(\frac{1}{2}g^2{n}^2t^2[{x},{p}])$. It is clear that the operation shifts the phase of the field depending on the commutator value, which is amplified by the number of photons in the optical field. If the commutation relation is modified, the small modification is amplified by the strength of the optical field. Another strong point is that the accuracy of the commutator measurement does not depend on the initial state of the oscillator.

The current values of experimental parameters allow the measurement of the commutator value upto $\beta_0\approx 10^{12}$ which is still far above the Planck length scale of $\beta_0\approx 1$ but still 22 orders of magnitude better than current large scale experiments. With a larger number of photons in the optical field and lower oscillator frequency, we can achieve an accuracy near to the Planck scale. However, the low frequency will cause fast decoherence of the oscillator and the large photon number field may make it unstable. As the authors point out the proposal is vulnerable to the so-called soccerball effect \cite{Hossenfelder2014a}. The canonical commutation relation is proportional to the unit operator which in turn gives a non-trivial lower bound of the uncertainty principle which does not depend on the state of the physical system. However, the modification terms in the GUP nonlinearly depend on position or momentum operators which brings the lower bound dependent on the quantum state and size of the system. Depending on the effect of modification either to the centre of mass motion or each constituent particles, the values of the modification will differ. This is discussed in the supplementary material of \cite{Pikovski2012}. 

\section{Conclusions}
When the effects of gravity cannot be ignored, a number of new quantum features are predicted to emerge. The commonly accepted view is that this should happen on extremely small lengths (or equivalent, at high energies), with most candidate theories placing the relevant scale at around that given by the Planck units\footnote{Though we stress not all.}, eq. (\ref{Planckunits}). As we have seen, semi-classical arguments alone are often sufficient to motivate many of the most widely expected effects. However, we should make clear that these are not the only effects predicted by quantum gravity theories. This review has made no reference to the implications for high energy physics, and consequences for particle physics or cosmology have been entirely overlooked. The interested reader can find details in this direction in reference \cite{Amelino-Camelia2013a}.

Yet, despite the apparent inaccessibility of these energy scales, there is a growing expectation that developments in experimental physics may soon be sensitive enough to a few of these features. Notably, constraints are already being placed on phenomenological models of the energy-momentum dispersion relations from a variety of sources, with new proposals potentially on the horizon. The prospects of a concrete discovery are exciting, however one must also be cautious. A rigorous connection between an explicit model (if one is even available) and measurement is difficult, and in general the features discussed here are only accessible after some kind of amplification over large distance or mass scales. This in turn introduces additional ambiguity \cite{Amelino-Camelia2013,Hossenfelder2014a}. Nevertheless, obtaining concrete measurements would have important consequences for our understanding of gravity, and will ultimately be essential in finally deciding the correct approach to take.  

\section{Acknowledgments}
This work was supported by a Leverhulme Trust Research Grant (Project RPG-2014-055), the UK EPRSC (grant number EP/J014664/1) and the  People  Programme (Marie Curie Actions) of the European Union's Seventh Framework Programme (FP7/2007-2013) under REA grant agreement n\textsuperscript{o}
317232.
\bibliographystyle{tandfx}
\bibliography{mybib}{}

\end{document}